\documentclass{amsart}
\usepackage{graphicx}
\usepackage{amssymb, amsmath}
\usepackage{amsfonts}
\usepackage{amssymb}
\usepackage{epsfig}
\usepackage{float}

\newtheorem{theorem}{Theorem}

\newtheorem{definition}[theorem]{Definition}

\newtheorem{lemma}[theorem]{Lemma}

\newtheorem{remark}[theorem]{Remark}

\newcommand{\divv}{\text{\rm div}}

\newcommand{\C}{\mathbb C}

\newcommand{\R}{\mathbb R}

\def\be{\begin{equation}}
\def\ee{\end{equation}}

\def\R{\mathbb R}

\def\tilde{\widetilde}

\numberwithin{equation}{section}
\numberwithin{theorem}{section}
\numberwithin{figure}{section}

\begin{document}
\bibliographystyle{siam}

\title[Double-diffusive Convection]
{Attractor Bifurcation of Three-Dimensional Double-Diffusive Convection}

\author[C. Hsia]{Chun-Hsiung Hsia}
\address[CH]{Department of Mathematics, Indiana University,
Bloomington, IN 47405}
\email{chsia@indiana.edu}

\author[T. Ma]{Tian Ma}
\address[TM]{Department of Mathematics, Sichuan University,
Chengdu, P. R. China}

\author[S. Wang]{Shouhong Wang}
\address[SW]{Department of Mathematics,
Indiana University, Bloomington, IN 47405}
\email{showang@indiana.edu}

\date{\today}

\thanks{The work was supported in part by the
Office of Naval Research,  by the National Science Foundation, 
and by the National Science Foundation
of China.}


\begin{abstract}
In this article, we present a bifurcation analysis on 
the  double-diffusive convection. Two pattern selections, rectangles and
squares, are investigated. It is proved that there
 are two different types of attractor bifurcations depending on
 the thermal and salinity Rayleigh numbers for each pattern. The analysis is 
based on a newly developed attractor bifurcation theory, together 
with eigen-analysis and the center manifold reductions.
\end{abstract}

\maketitle
\section{Introduction}

The main objective of this article is to develop a 
bifurcation and stability theory for the double-diffusive convection, 
using a newly developed bifurcation theory for dynamical systems, both finite 
and infinite dimensional \cite{b-book}. 
Convective motions occur in a fluid when there are density 
variations present. Double-diffusive convection is the name 
given to such convective motions when the density variations 
are caused by two different components which have different 
rates of diffusion. 
Double-diffusion was first originally discovered 
in the 1857 by Jevons \cite{jevons}, forgotten, and then rediscovered
as an ``oceanographic curiosity'' a century later; see  among others 
Stommel, Arons and 
Blanchard \cite{stommel}, Veronis \cite{gv}, 
and Baines and Gill \cite{pg}. 

We have conducted a bifurcation and stability analysis in \cite{hmw} for the 
{\it two dimensional} (2D) double-diffusive convection model. In this article, 
we continue our analysis to study the {\it three dimensional} (3D)
double-diffusive convection problem. As in the two-dimensional case, 
the governing equations are  the Boussinesq equations with 
two diffusion equations of the temperature and salinity functions.

The double-diffusive system involves four important nondimensional 
parameters:  the thermal Rayleigh number
$\lambda$, the solute Rayleigh number $\eta$, the Prandtl number $\sigma$
and the Lewis number $\tau$. 
We examine in this article different transition/instability regimes defined by 
these parameters. We study the onset of double 
diffusive instabilities in these regimes.
In the ocean circulation case, the Prandtl and Lewis numbers satisfy  
$\sigma>1$  and $\tau <1$, as the
heat diffuses about 100 times more rapidly than salt \cite{stern}. 
In this case, different regimes of stabilities and instabilities/transitions 
of the basic state can be described by regions in the $\lambda$-$\eta$ plane
(the thermal and salt Rayleigh numbers) as shown in 
Figure~\ref{fig4.1}. In this article, we focus on the regimes where 
$ \eta <  \eta_c$; see Figure~\ref{fig4.1}.
In the case where $\eta >  \eta_c$, transitions to periodic or aperiodic 
solutions are expected, and will be addressed elsewhere.

\medskip
In both 2D and 3D cases, the central gravity of the analysis
is the reduction of the problem 
to the center manifold in the first unstable eigendirections, 
based on an approximation formula for the center manifold function. 
The key idea is to find  the  approximation of the 
reduction to certain order, leading to a  ``nondegenerate'' system 
with higher order perturbations.
The analysis for the 3D case 
studied here has all the difficulties appeared 
in the analysis for the 2D case such as 
the nonsymmetric linearized eigenvalue problem. Additional difficulties 
occur in the 3D case. One difficulty is the lack of the existence of 
global strong solutions of the 3D Boussinesq equations, and another is 
the higher dimensionality of the generalized eigenspace, leading to 
the center manifold reduction to higher dimensional center manifolds, and 
consequently to more complicated dynamics.


\medskip

With this reduction in our disposal, a general 
bifurcation theorem follows from the general strategy of 
attractor bifurcations. We prove in
particular that there are two different types of 
transition/bifurcation from the basic state: one is subcritical and 
the other is supercritical. In the rectangle case (i.e. the horizontal 
domain is a rectangle), the types of transition/bifurcation are 
dictated explicitly by a nondimensional parameter $\eta_{c_1}$; see 
Figure~\ref{fig4.1}. In the square case (i.e. the horizontal 
domain is a square), types are determined by the 
quadratic form (\ref{eq2.12}).
From the physical point of view, the supercritical bifurcation corresponds to 
a continuous transition. The subcritical bifurcation corresponds, however,  
to a jump transition, a very different transition. As demonstrated 
in the 2D case \cite{hmw}, this subcritical bifurcation leads to the 
existence of  a saddle-node bifurcation and the hysteresis phenomena, 
thanks to the existence of strong solutions and the existence 
of global attractors. In the 3D case studied here in this article, 
we are not able to make such conclusions. However, we can  still prove that 
there is an absorbing region resembling 
the main properties of the saddle-node bifurcation and hysteresis.

\medskip

There have been extensive studies on bifurcation and stability 
analysis for convection models; 
see among others \cite{wn,busse,bg1,bg2}  and the references therein.
We mention in particular that 
in \cite{bg1}, Buzano and Golubitsky used the singularity-theory method 
to study the problem of pattern formation as it relates to bifurcation
with respect to the hexagonal lattice. They treated bifurcation problems 
that are symmetric with respect to group preserving doubly periodic functions 
on a hexagonal lattice. Their method provides a delicate alternative 
to study the convection problem in this paper.

Another remark is that as mentioned by Sattinger 
on page 96 in \cite{sattinger},
a stable bifurcating solution with respect to a subclass of disturbances
of full symmetry might turn out to be unstable with respect to
the disturbances of full symmetry. In our analysis, although we use the 
symmetry properties implicitly to reduce the calculations, 
we don't use the equivariant theory to reduce 
the problem to a lower dimensional space. Since we  want to keep  
as much stability information as possible. 
  
This article is organized as follows.
The basic governing equations are given in Section 2. 
The main theorems are stated in Section 3. The remaining sections 
are devoted to the proof of the main theorems, with
 Section 4 on eigenvalue problems, Section 
5 on center manifold reductions and the proofs of the theorems.

\section{The Double-Diffusive Equations}
The nondimensional  double-diffusive convection 
problem in  a three-dimensional (3D) domain  $\Omega=\mathbb R^2 \times 
(0, 1) \subset \R^3$ with coordinates denoted by $(x, y, z)$ are 
given as follows; see  Veronis \cite{gv}: 
\be
\label{eq2.4}
\left\{
\begin{aligned}
& \frac{\partial U}{\partial t}=\sigma( \Delta U-\nabla p)
  +\sigma(\lambda T-\eta S)e-(U \cdot \nabla)U,\\
& \frac{\partial T}{\partial t}=\Delta T +w-(U \cdot \nabla)T,\\
& \frac{\partial S}{\partial t}=\tau \Delta S+w- (U \cdot \nabla)S,\\
& \divv  U = 0,
\end{aligned}
\right.
\ee
where $U=(u,v,w)$, $\lambda$ the thermal Rayleigh number, $\eta$ the 
salinity Rayleigh number, $\sigma$ the  Prandtl number, and $\tau$ 
the Lewis number. We consider the periodic boundary condition 
in the $x$ and $y$ directions 
\begin{align}
\label{eq2.5}
 (U,T,S)(x,y,z,t)& =(U,T,S)(x+2j\pi/\alpha_1,y,z,t) \\
&   =(U,T,S)(x,y+2k\pi/\alpha_2,z,t), \nonumber
\end{align}
for any $j,k\in \mathbb Z$. At the top and bottom boundaries, we impose 
 the free-free boundary conditions; namely, 
\be
\label{eq2.6}
(T, S, w)=0, \quad \frac{\partial u}{\partial z}=0,
\quad \frac{\partial v}{\partial z}=0, \quad \text{at} \quad z=0,1. 
\ee 
It's natural to put the constraint
\begin{equation}
\label{eq2.7}
\int_{\Omega} udxdydz=\int_{\Omega}vdxdydz=0
\end{equation}
for the  problem (\ref{eq2.4})-(\ref{eq2.6}).
It is easy to see that (\ref{eq2.4})
is invariant under this constraint.
The initial value conditions are given by
\be
\label{eq2.10}
(U,T,S)=(\tilde{U},\tilde{T},\tilde{S}) \quad  \text{at}\quad t=0.
\ee


Let 
\begin{align*}
 H=& \{(U,T,S)\in L^{2}(\Omega)^{5}\mid \divv \, U=0, w\mid_{z=0,1}=0, 
      (u,v) \text{ satisfy}\,\, (\ref{eq2.5})\,\,
    \text{and}\,\, (\ref{eq2.7})\},\\
 H_1=& \{(U,T,S)\in H^{2}(\Omega)^{5}\cap H \,|\,(U,T,S)\,\text{ satisfies }
  \, (\ref{eq2.5}) \text{ and } \,(\ref{eq2.6})  \}.
\end{align*}
Let $G: H_{1} \to H$, and 
$L_{\lambda \eta}=-A-B_{\lambda \eta}: H_{1} \to H$ be defined by
\begin{align*}
& G (\psi) = (- P[(U\cdot\nabla )U], 
                       - (U\cdot \nabla )T, 
                       - (U\cdot \nabla )S ),  \\
& A\psi = ( -P [\sigma(\Delta U)], -\Delta T, -\tau \Delta S ), \\
& B_{\lambda \eta} \psi =  (-P [\sigma(\lambda T-\eta S)e],-w,-w ), 
\end{align*} 
for any $\psi=(U, T, S) \in H_1$. Here $P$ is the Leray projection to 
$L^2$ fields, and for a detailed account of the function spaces, 
see among many others \cite{temam}. 
Then the Boussinesq equations (\ref{eq2.4})-(\ref{eq2.7}) can be written 
in the following operator form
\begin{equation}\label{eq2.11}
\frac{d\psi}{dt} = L_{\lambda \eta} \psi + G(\psi), \qquad \psi=(U,T,S).
\end{equation}

It is classical that there is a global weak solution for the system, and 
there is a global strong solution for small data; see among 
others \cite{fmt, LTW92b}. Of course, the global existence of strong 
solutions is not known for large data.

\section{Main Results}  
\subsection{Attractor bifurcation theory}

In this subsection, we recapitulate the attractor bifurcation theory
introduced by two of the authors in \cite{mw-db1,b-book}.

Let  $H$ and  $H_1$ be two Hilbert spaces,
and $H_1 \hookrightarrow H$ be a dense and compact inclusion.
We consider the following
nonlinear evolution equations
\be
\label{eq3.1}
\left\{
\begin{aligned}
& \frac{du}{dt} = L_\lambda u +G(u,\lambda), \\
& u(0) = u_0,
\end{aligned}
\right.
\ee
where $u: [0, \infty) \to H$  is the unknown function, $\lambda \in
\mathbb R$  is the  system  parameter, and
$L_\lambda:H_1\to H$ are parameterized linear completely
continuous fields depending continuously on $\lambda\in \R^1$, which
satisfy
\begin{equation}
\label{eq3.2}
\left\{\begin{aligned}
& -L_\lambda = A + B_\lambda && \text{a sectorial operator}, \\
& A:H_1 \to H && \text{a linear homeomorphism}, \\
& B_\lambda :H_1\to H && \text{parameterized linear compact
operators.}
\end{aligned}\right.
\end{equation}
It is easy to see \cite{henry} that $L_\lambda$
generates an analytic semi-group $\{e^{tL_\lambda}\}_{t\ge 0}$.
Then we can define  fractional power operators $(-L_\lambda)^{\mu}$ for any
$0\le \mu \le 1$ with domain $H_\mu = D((-L_\lambda)^{\mu})$ such that
$H_{\mu_1} \subset H_{\mu_2}$ if $\mu_1 > \mu_2$, and $H_0=H$.

Furthermore, we assume that the nonlinear terms
$G(\cdot, \lambda):H_\mu \to H$ for some $1> \mu \ge 0$
are a family of parameterized $C^r$
bounded operators ($r\ge 1$) continuously depending on the parameter
$\lambda\in \R^1$, such that
\begin{equation}
\label{eq3.3}
 G(u,\lambda) = o(\|u\|_{H_\mu}), \quad \forall\,\, \lambda\in \R^1.
\end{equation}

In this paper, we are interested in the sectorial operator
$-L_\lambda = A +B_\lambda$ such
 that there exist an eigenvalue sequence $\{\rho_k\}
\subset \C^1$ and an eigenvector sequence $\{e_k, h_k\}\subset
H_1$ of $A$:
\begin{equation}
\label{eq3.4}
\left\{\begin{aligned}
& Az_k = \rho_kz_k,  \qquad z_k=e_k + i h_k, \\
& \text{Re} \rho_k\to \infty \,\,(k\to\infty), \\
& |\text{Im} \rho_k / (a + \text{Re} \rho_k) | \le c,
\end{aligned}\right.
\end{equation}
for some $a, c > 0$, such that
$\{e_k, h_k\}$ is a basis of $H$.
Also we assume that
there is a constant $0<\theta<1$ such that
\begin{equation}
\label{eq3.5}
B_\lambda :H_\theta \longrightarrow H \,\,\text{bounded, $\forall$
$\lambda\in \R^1$.}
\end{equation}
Under conditions (\ref{eq3.4}) and (\ref{eq3.5}), the operator
$-L_\lambda=A + B_\lambda$ is a sectorial operator.

Let $\{S_\lambda(t)\}_{t\ge 0}$ be an operator semi-group generated by
the equation (\ref{eq3.1}).
Then the solution of (\ref{eq3.1})  can be expressed
as $\psi(t, \psi_0) = S_\lambda(t)\psi_0, \qquad t\ge 0.$

Consider (\ref{eq3.1}) satisfying (\ref{eq3.2})  and (\ref{eq3.3}). 
We start with the Principle of Exchange of Stabilities (PES).
Let the eigenvalues (counting the multiplicity) of $L_\lambda$ be given by
$\beta_1(\lambda)$, $\beta_2(\lambda)$, $\cdots$.
Suppose that
\begin{equation}
\label{eq3.13} Re\beta_i(\lambda)\begin{cases}
<0, \,\,\,\, \text{if}\,\,\,\, \lambda<\lambda_0\\
=0, \,\,\,\, \text{if}\,\,\,\, \lambda=\lambda_0\\
>0, \,\,\,\, \text{if}\,\,\,\, \lambda>\lambda_0
\end{cases}
\qquad(1\le i\le m)
\end{equation}

\begin{equation}
\label{eq3.14} Re\beta_j(\lambda_0)<0.\qquad \forall \,\,m+1\le j.
\end{equation}

Let the eigenspace of $L_\lambda$ at $\lambda_0$ be
\[
E_0=\displaystyle{\bigcup_{1\le j \le m}\bigcup_{k=1}^{\infty}}\{u,v\in H_1\mid
(L_{\lambda_0}-\beta_j(\lambda_0))^k w=0, w=u+iv \}.
\]
It is known that $\dim E_0=m$.

\begin{theorem}[T. Ma and S. Wang \cite{mw-db1,b-book}]
\label{th3.6} 
Assume that the conditions (\ref{eq3.2})-(\ref{eq3.5}) and
(\ref{eq3.13})-(\ref{eq3.14}) hold true, 
and $u=0$ is locally asymptotically 
stable  for (\ref{eq3.1}) 
at $\lambda=\lambda_0$. Then the following assertions hold true.

\begin{enumerate}
\item (\ref{eq3.1}) bifurcates from $(u,\lambda)=(0,\lambda_0)$ to 
attractors $\Sigma_\lambda$, having the same homology as  $S^{m-1}$, 
for $\lambda>\lambda_0$, 
with $ m-1\le dim\Sigma_\lambda\le m$, which is connected 
if $ m>1$;

\item For any $u_\lambda\in\Sigma_\lambda$, $u_\lambda$ can 
be expressed as
\[
u_\lambda=v_\lambda+o(\|v_\lambda\|_{H_1}), \,\, v_\lambda\in E_0;
\]
\item There is an open set $U\subset H$ with $0\in U$ such that the 
attractor $\Sigma_\lambda$ bifurcated from $(0,\lambda_0)$ 
attracts $U\backslash\Gamma$ in $ H$, where $\Gamma$ is the stable 
manifold of $u=0$ with co-dimension m.
\end{enumerate}
\end{theorem}



\subsection {Main theorems}
We now consider the double diffusive convection equations (\ref{eq2.4}). 
In this article, we always consider the case 
where the parameters $\lambda$ 
and $\eta$ satisfy
\begin{equation}
\label{eq2.9}
\begin{aligned}
 & \eta < \eta_c=\frac{27}{4}\pi^4\tau^2(1+\sigma^{-1})(1-\tau)^{-1},\\
 & \lambda \approx \lambda_c= \frac{\eta}{\tau}+\frac{27}{4}\pi^4.
\end{aligned}     
\end{equation}
Two pattern selections, rectangles and squares, are studied.

\subsubsection{Rectangle Case}
For the rectangle case, we assume that
\begin{align*}
 & \alpha_1=\frac{\pi}{\sqrt{2}}  \qquad \text{and} 
\qquad \alpha_2 \ne \alpha_1.   \end{align*}
Here the condition on $\alpha_1$ and $\alpha_2$ 
defines the aspect ratios of the domain. The first eigenspace 
of the eigenvalue problem (\ref{eq4.1}) is of dimension two.
\begin{figure}
 \centering \includegraphics[height=.3\hsize]{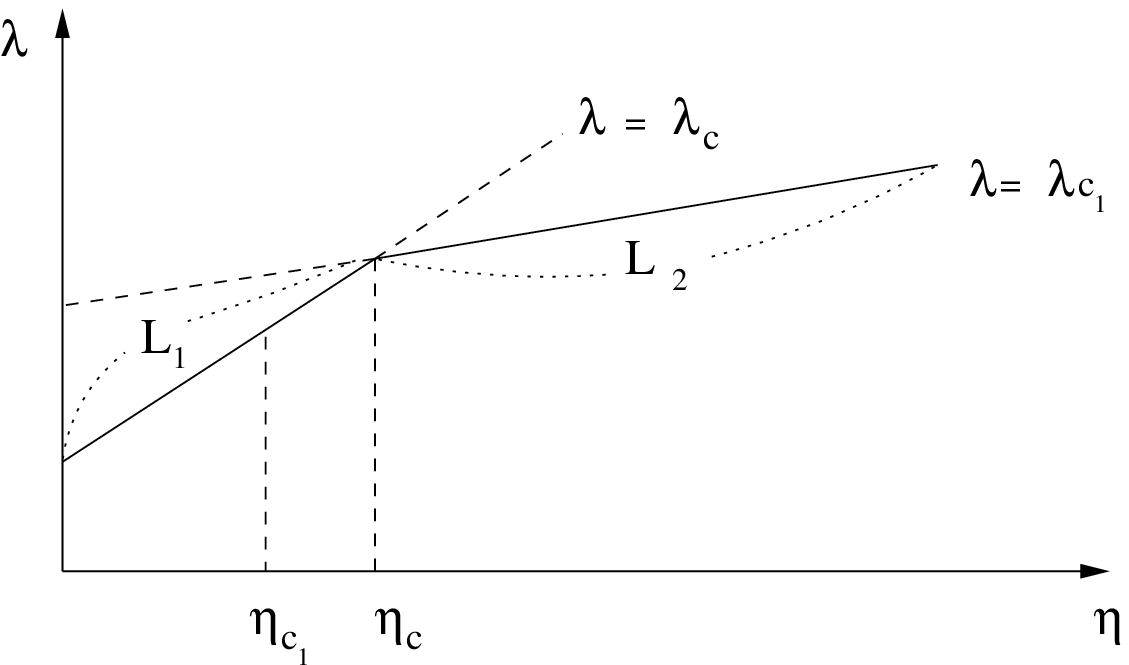}
 \caption{}
\label{fig4.1}
\end{figure}
First we consider a more physically relevant diffusive regime  
where the thermal Prandtl number $\sigma$ 
is bigger than $1$, and the Lewis number $\tau$ is less than $1$:
\begin{equation}
\label{eq2.8}
 \sigma> 1 >\tau. 
\end{equation}
In this case, we consider two straight lines in the 
$\lambda-\eta$ parameter plane as shown in Figure~\ref{fig4.1}:
\begin{equation}
\left\{
\begin{aligned}
& L_1: \quad \lambda=\lambda_c(\eta) 
   =\frac{\eta}{\tau}+\frac{27}{4}\pi^4, \\
& L_2: \quad \lambda=\lambda_{c_1}(\eta)= \frac{(\sigma+\tau)}{(\sigma+1)}\eta+
      \frac{27}{4}\pi^4(1+\sigma^{-1}\tau)(1+\tau).
\end{aligned}
\right. \label{cc}
\end{equation}
Also shown in Figure~\ref{fig4.1} are two critical values for $\eta$ given by 
$$\eta_c=\frac{27}{4}\pi^4\tau^2(1+\sigma^{-1})(1-\tau)^{-1}, \quad 
\eta_{c_1}=\frac{27}{4}\pi^4\tau^3(1-\tau^2)^{-1}.$$

The following two theorems study the transitions/bifurcation 
of the double-diffusive model near the line $L_1$ for 
$\eta < \eta_c$.
\begin{theorem}
\label{th2.3}
Assume that the condition (\ref{eq2.9}) holds true, $\sigma>1>\tau$, and
$\eta<\eta_{c_1}=\frac{27}{4}\pi^4\tau^3(1-\tau^2)^{-1}$. Then the following 
assertions for the problem (\ref{eq2.4})-(\ref{eq2.7})  hold true.
\begin{enumerate}
\item If $\lambda \le \lambda_c$, the steady state $(U,T,S)=0$ is
      locally asymptotically stable for the problem.
\item For $\lambda>\lambda_c$, the solutions bifurcate 
from $((U,T,S),\lambda)=(0,\lambda_c)$ to 
an attractor $\Sigma_\lambda=S^{1}$,  
consisting of steady state solutions  
of the problem.  
 \end{enumerate} 
\end{theorem}
\begin{figure}
 \centering \includegraphics[height=.4\hsize]{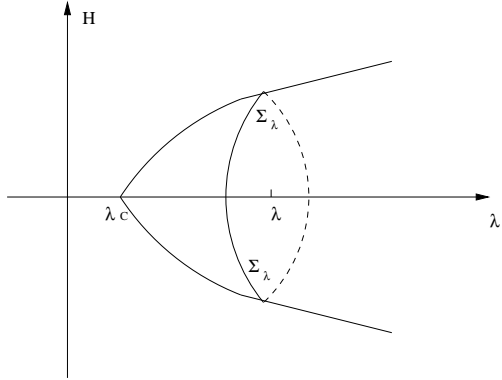}
 \caption{If $\tau>1$ or $\eta<\eta_{c_1}$, the solutions
          bifurcate from $(0,\lambda_c)$ to an attractor 
         $\Sigma_{\lambda}$ for $\lambda>\lambda_c$. }
 \label{fig2.1}
\end{figure}

\begin{theorem}
\label{th2.4}
Assume that the condition (\ref{eq2.9}) holds true, $\sigma>1>\tau$ and
$\eta_c>\eta>\eta_{c_1}=\frac{27}{4}\pi^4\tau^3(1-\tau^2)^{-1}$. 
Then on $\lambda < \lambda_c$,  
the problem (\ref{eq2.4})-(\ref{eq2.7}) bifurcates 
from $((U,T,S),\lambda)=(0,\lambda_c)$ 
      to a repeller  $\Sigma_{\lambda}=S^1$, consisting of 
steady state solutions of the problem.
\end{theorem}
\begin{figure}
 \centering \includegraphics[height=.4\hsize]{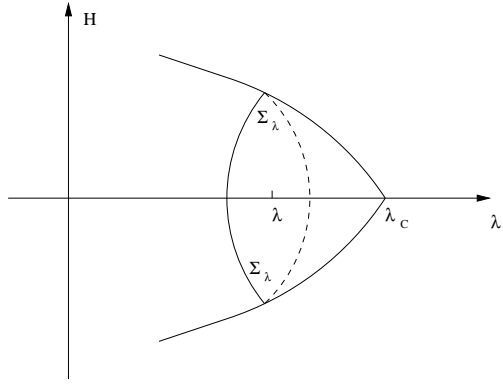}
 \caption{If $\tau<1$ and $\eta_{c_1}<\eta<\eta_c$, the problem
        has a subcritical type bifurcation for $\lambda<\lambda_c$.}
 \label{fg2.2}
\end{figure}

\begin{figure}
 \centering \includegraphics[height=.3\hsize]{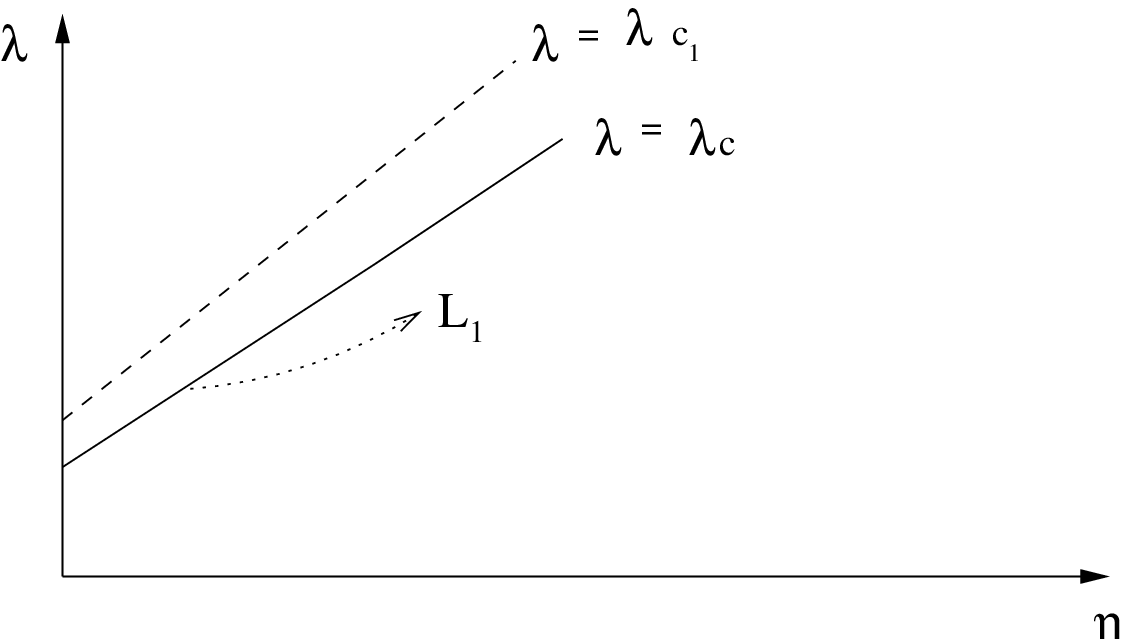}
 \caption{}
 \label{fig4.2}
\end{figure}

We now consider the diffusive parameter regime where 
$\sigma>1$, $\tau>1$,  and $\sigma\ne \tau$. In this case, 
two lines are shown in Figure~\ref{fig4.2}.
The following theorem provides bifurcation when $\lambda$ 
crosses the line $L_1$.

\begin{theorem}
\label{th2.5}
Assume that  $\sigma>1$, $\tau>1$, $\sigma\ne \tau$  and
 (\ref{eq2.9}) hold true, then for any $\eta>0$, the following assertions 
for the problem  (\ref{eq2.4})-(\ref{eq2.7})  hold true.
\begin{enumerate}
\item If $\lambda \le \lambda_c$, the steady state $(U,T,S)=0$ is
      locally asymptotically stable for the problem.
\item For $\lambda>\lambda_c$, the solutions bifurcate 
from $((U,T,S),\lambda)=(0,\lambda_c)$ to 
an attractor $\Sigma_\lambda$,
homeomorphic to $S^{1}$, which consists of steady state solutions  
of the problem.  
 \end{enumerate} 
\end{theorem}

\medskip

\subsubsection{Square Case}

For the square case, we assume $\alpha_1=\alpha_2=\frac{\pi}{2}$. 
Then, the first eigenspace of the eigenvalue
problem (\ref{eq4.1}) is of dimension 4. In this case, the bifurcation
type is determined by 
the following  quadratic form
\begin{equation}
\label{eq2.12}
\delta_1(x_{1111}^2+y_{1111}^2+x_{1-111}^2+y_{1-111}^2)^2
+2\delta_2 (x_{1111}^2+y_{1111}^2)(x_{1-111}^2+y_{1-111}^2),
\end{equation}
 where
$\delta_1$ and $\delta_2$ are as  defined in (\ref{eq5.11})
and (\ref{eq5.12}), which can
 be evaluated 
precisely when the parameters $\tau$, $\sigma$, $\eta$ and
 $\lambda$ are given. More precisely, we have the following 
two theorems.

\begin{theorem}
\label{th2.6}
Assume that  (\ref{eq2.9}) holds true and 
(\ref{eq2.12}) is negative definite, then the following assertions 
for the problems (\ref{eq2.4})-(\ref{eq2.7}) hold true.
\begin{enumerate}
\item If $\lambda \le \lambda_c$, the steady state $(U,T,S)=0$ is
      locally asymptotically stable for the problem.
\item For $\lambda>\lambda_c$, the solutions 
bifurcate from $((U,T,S),\lambda)=(0,\lambda_c)$ to 
an attractor $\Sigma_\lambda$, having the same homology as $S^{3}$.
\end{enumerate}
\end{theorem}

\begin{theorem}
\label{th2.7}
Assume that  (\ref{eq2.9}) holds true and 
(\ref{eq2.12}) is positive definite. Then on $\lambda < \lambda_c$, 
the problem  (\ref{eq2.4})-(\ref{eq2.7}) bifurcates from 
$((U,T,S),\lambda)=(0,\lambda_c)$ 
to a repeller $\Sigma_{\lambda}$, 
having the same homology as  $S^3$.
\end{theorem}
\section{Eigenvalue problem}
In order to apply the center manifold theory to reduce the 
bifurcation problems, we consider the following eigenvalue 
problem for the linearized equations of (\ref{eq2.4})-
(\ref{eq2.7}):
\be
\label{eq4.1}
\left\{
\begin{aligned}
& \sigma(\Delta U- \nabla p)+\sigma( \lambda T- \eta S)e=\beta U,\\
& \Delta T + w =\beta T,\\
& \tau \Delta S + w = \beta S,\\
& \divv U = 0,
\end{aligned}
\right.
\ee 
supplemented with (\ref{eq2.6}) and (\ref{eq2.7}).

\subsection{Eigenvalues}
We shall use the method of  separation of variables to deal with
the problem (\ref{eq4.1}). Since $\psi=(U,T,S)$ satisfies the
periodic condition (\ref{eq2.5}), we expand the fields in 
Fourier series as
\be
\label{eq4.2}
\psi(x,y,z)=\sum_{j,k=-\infty}^{\infty}\psi_{jk}(z)e^{i(j\alpha_1 x
+k \alpha_2 y)}.
\ee 

Plugging (\ref{eq4.2}) into (\ref{eq4.1}), we obtain the following 
ODE system:
\be
\label{eq4.3}
\left\{
\begin{aligned}
& D_{jk} u_{jk} - ij\alpha_1 p_{jk}=\sigma^{-1}\beta u_{jk},\\
& D_{jk} v_{jk} - ik\alpha_2 p_{jk}=\sigma^{-1}\beta v_{jk},\\
& D_{jk} w_{jk}-p_{jk}'+\lambda T_{jk} - \eta S_{jk}=
   \sigma^{-1}\beta w_{jk},\\
& D_{jk} T_{jk} + w_{jk} = \beta T_{jk},\\
& \tau D_{jk} S_{jk} + w_{jk} = \beta S_{jk}, \\
& ij\alpha_1 u_{jk}+ik\alpha_2 v_{jk} + w_{jk}'=0,\\
& u_{jk}' \mid_{z=0,1} = v_{jk}' \mid_{z=0,1}= 
w_{jk}\mid_{z=0,1}= T_{jk}\mid_{z=0,1}= S_{jk}\mid_{z=0,1}=0, 
\end{aligned} 
\right.
\ee
for $j, k \in \mathbb Z$, where $'=d/dz$, $D_{jk}=d^2/dz^2-\alpha_{jk}^2$
and $\alpha_{jk}^2=j^2\alpha_1^2+k^2\alpha_2^2$. If $w_{jk}\ne 0$,   
(\ref{eq4.3}) can be reduced to  a single equation for $w_{jk}(z)$:
\begin{align}
& \label{eq4.4} \{(\tau D_{jk}-\beta)
  (D_{jk}-\beta)(D_{jk}-\sigma^{-1}\beta)D_{jk}\\
& \qquad  +\alpha_{jk}^2[\lambda(\tau D_{jk}-
     \beta)-\eta(D_{jk}-\beta)]\}w_{jk}=0,
\nonumber \\
&\label{eq4.5}
 w_{jk} = w_{jk}'' = w_{jk}^{(4)} 
= w_{jk}^{(6)} = 0  \quad \text{at} \quad z=0,1,
\end{align}
for $j,k \in \mathbb Z$. Thanks to ({\ref{eq4.5}}), $w_{jk}$
can be expanded in a  Fourier sine series
\be
\label{eq4.6}
w_{jk}(z)=\sum_{l=1}^{\infty}w_{jkl}\sin l\pi z,
\ee
for $(j,k)\in \mathbb Z \times \mathbb Z$. 
Substituting (\ref{eq4.6}) into (\ref{eq4.4}),
we see that the corresponding
 eigenvalues $\beta$ of Problem (\ref{eq4.1}) satisfy
the cubic equations
\begin{align}
& \label{eq4.7}
\beta^3+(\sigma+\tau+1)\gamma_{jkl}^2\beta^2+[(\sigma+\tau+\sigma \tau)
\gamma_{jkl}^4-\sigma \alpha_{jk}^2\gamma_{jkl}^{-2}(\lambda-\eta)]\beta\\
& \qquad \qquad \qquad
+\sigma \tau \gamma_{jkl}^6 + \sigma \alpha_{jk}^2 (\eta - \tau \lambda)
=0, \nonumber
\end{align}
for $j,k\in \mathbb Z$ and $l\in \mathbb N$,
where $\gamma_{jkl}^2=\alpha_{jk}^2 + l^2\pi^2$.

For the sake of convenience to analyze the distribution of the eigenvalues, 
we now introduce some notations. 
For fixed parameters $\sigma$, $\tau$, $\eta$ and $\lambda$, let
\begin{align*}
& g_{jkl}(\beta)=\beta^3+(\sigma+\tau+1)\gamma_{jkl}^2\beta^2+
(\sigma+\tau+\sigma \tau)\gamma_{jkl}^4\beta+\sigma \tau \gamma_{jkl}^6, \\
& h_{jkl}(\beta)=[\sigma \alpha_{jk}^2\gamma_{jkl}^{-2}
    (\lambda-\eta)]\beta
-\sigma \alpha_{jk}^2 (\eta - \tau \lambda), \\
& f_{jkl}(\beta)=g_{jkl}(\beta)-h_{jkl}(\beta), \qquad
 \eta_{c}=\frac{27}{4}\pi^4\tau^2(1+\sigma^{-1})(1-\tau)^{-1}, \\
& \eta_{c_1}=\frac{27}{4}\pi^4\tau^3(1-\tau^2)^{-1}, \qquad
 \lambda_{c}=\frac{\eta}{\tau}+\frac{27}{4}\pi^4.
\end{align*}
Furthermore, let $\beta_{jkl1}$, $\beta_{jkl2}$ and $\beta_{jkl3}$ 
be the zeros of $f_{jkl}$ with 
$$ Re(\beta_{jkl1})\geq Re(\beta_{jkl2}) 
\geq Re(\beta_{jkl3}), $$ and  let 
$\beta_{jkl \sigma}=-\sigma \gamma_{jkl}^2$.

The following lemma characterizes the eigenvectors 
of $L_{\lambda \eta}$ with zero $w$-components; 
the proof is straightforward and we omit the details.
   
\begin{lemma}
\label{le4.2}
For $l\in 0\cup \mathbb N$ and $(j,k)\in \mathbb Z^2$,
$\beta_{jkl \sigma}$ is an eigenvalue of the problem (\ref{eq4.1}). 
Moreover, the following assertions are true:

\begin{enumerate}
\item If $(j,k)=(0,0)$ and $l> 0$, the corresponding
       eigenvectors  are
     \begin{equation}
     \label{eq4.8}
      \psi^{\beta_{00l \sigma}}_1= ( \cos l\pi z, 0, 0, 0, 0)^{t}, 
      \qquad  \psi^{\beta_{00l \sigma}}_2=( 0, \cos l\pi z, 0, 0, 0)^{t};
       \end{equation} 
     
\item If $j^2+k^2\ne 0$ and $l\in 0\cup \mathbb N$, the corresponding
      eigenvectors are
      \begin{equation}
      \label{eq4.9}
    \begin{aligned}
  &  \psi^{\beta_{jkl \sigma}}_1=
        ( k \cos(j\alpha_1 x + k\alpha_2 y) \cos l\pi z,
               -j \cos(j\alpha_1 x + k\alpha_2 y) \cos l\pi z, 0, 0, 0)^{t},\\
  & \psi^{\beta_{jkl \sigma}}_2=
              ( k \sin (j\alpha_1 x + k\alpha_2 y) \cos l\pi z,      
               -j \sin (j\alpha_1 x + k\alpha_2 y) \cos l\pi z, 0, 0, 0)^{t}.             \end{aligned}
\end{equation}
\end{enumerate}
\end{lemma}

In the following discussions, we shall focus on the following 
diffusive regime:
\be
\label{eq4.10}
\quad \sigma>1>\tau>0,\,\, \eta<\eta_{c}\,\, 
\text{and}\,\, \lambda\approx \lambda_c. 
\ee

\begin{lemma}
\label{le4.3}
 Suppose that $\alpha_{jk}^2=\frac{\pi^2}{2}$ 
 and (\ref{eq4.10}) holds true, 
    then  $f_{jk1}(\beta)$ has three 
      simple real zeros.
\end{lemma} 
\begin{proof}
Since $\lambda \approx \lambda_c$, it suffices to prove this statement for
 $\lambda=\lambda_c$. In this case,
\begin{align*}
 f_{jk1}(\beta)=& \beta^3+[3\pi^2(\sigma + \tau +1)/2]\beta^2
 +[9\pi^4(\sigma + \tau + \sigma \tau)/4 - \sigma(\lambda_c -\eta)/3] \beta\\
              =& f(\beta) \beta.
\end{align*}
Since $\eta < \eta_c$, the constant term of $f(\beta)$ is positive.
Hence $\beta_{jk11}=0$ is a simple zero of $f_{jk1}$.
Moreover, the quadratic discriminant of 
$f(\beta)$ is $9\pi^4 (\sigma + 1 - \tau)^2/4 + 
4 \sigma \eta (1- \tau)/(3 \tau) >0 $ . 
This implies $f_{jk1}$ has three 
simple real zeros.
\end{proof}

 We summarize the following important lemma about the distribution of 
 the zeros of $f_{jkl}$. From the physical point of view, 
this lemma verifies the principle of exchange of stabilities (PES).

\begin{lemma}
\label{le4.4}
 Assume that either \,  1) $0<\eta<\eta_c$ with $\tau<1$ \,\, or
   \,\, 2) $\eta>0$ with $\tau>1$ \, holds true, then
\begin{align}
& 
\label{eq4.11}
\beta_{jk11}(\lambda)\begin{cases}
<0 \,\,\,\, \text{if} \,\,\,\, \lambda<\lambda_c\\
=0 \,\,\,\, \text{if} \,\,\,\, \lambda=\lambda_c\\
>0 \,\,\,\, \text{if} \,\,\,\, \lambda>\lambda_c
\end{cases}
&& \quad \text{if} \quad \alpha_{jk}^2=\frac{\pi^2}{2}, \\
&
\label{eq4.12}
Re\beta_{jklq}(\lambda)<0 &&\quad \text{if}\,\,\, 
 (\alpha_{jk}^2,l,q)\ne(\frac{\pi^2}{2},1,1).
\end{align}           
\end{lemma}

Before proceeding to the proof, we make a few remarks as follows.

\begin{remark}
\label{rm4.5}
{\rm  
The distribution of the zeros of $f_{jkl}$ was first partially  
analyzed by Veronis \cite{gv}; see also P.G. Baines \cite{pg}. 
The complete proof of this lemma is similar to the 2D case in \cite{hmw},
we shall omit it.
}
\end{remark}

 To check that the operators $-L_{\lambda \eta}$ 
satisfy condition (\ref{eq3.4}), we prove the following lemma.
\begin{lemma}
\label{le4.6}
\begin{enumerate}
\item All but finitely many zeros of  $f_{jkl}(\beta)$ 
      are negative real numbers for $(j,k,l)\in \mathbb Z^2 \times \mathbb N$.
\item $\beta_{jklq}\to-\infty$  if   $j^2+k^2+l^2\to \infty$.
\end{enumerate}
\end{lemma}
\begin{proof}
Since $f_{jkl}=g_{jkl}-h_{jkl}$, $\beta$ is 
a zero of $f_{jkl}(\beta)$ if and only
 if $\beta$ satisfies the equation
\begin{equation}
\label{eq4.15} 
g_{jkl}(\beta)=h_{jkl}(\beta).
\end{equation}
Plugging  $\beta=\gamma_{jkl}^{2}\beta^{*}$ into (\ref{eq4.15}), we obtain
\begin{equation}
\label{eq4.16}
(\beta^{*}+1)(\beta^{*}+\tau)(\beta^{*}+\sigma)=\vartheta_{jkl}[(\lambda-\eta)
\beta^{*}-(\eta-\tau\lambda)],
\end{equation} 
where $\vartheta_{jkl}=\sigma\alpha_{jk}^2/\gamma_{jkl}^{6}$. Since 
$lim_{j^2+k^2+l^2\to\infty}\vartheta_{jkl}=0$, the roots of (\ref{eq4.16})
 must be  negative 
real numbers near the interval $[-\sigma, -\tau]$ when $(j^2+k^2+l^2)$ 
is large. This completes the proof. 
\end{proof}
 

\subsection{Eigenvectors}
We now make some observations to analyze the spectrum of $L_{\lambda \eta}$. 
Since $g_{jkl}(\beta)=(\beta+\gamma_{jkl}^2)
(\beta+\tau\gamma_{jkl}^2)(\beta+\sigma\gamma_{jkl}^2)$ and $h_{jkl}=
\sigma \alpha_{jk}^2\gamma_{jkl}^{-2}[(\lambda-\eta)\beta
-(\eta - \tau \lambda)\gamma_{jkl}^2]$, it's easy to check that 
$\beta=-\gamma_{jkl}^2$ or $\beta=-\tau\gamma_{jkl}^2$ is a zero 
of $f_{jkl}(\beta)$ if and only if $\alpha_{jk}=0$. In the case 
of $\alpha_{jk}^2=0$, the
zeros of $f_{00l}$ are $-\tau\gamma_{00l}^2$ ($\beta_{00l1}$),
  $-\gamma_{00l}^2$ ($\beta_{00l2}$) and
$-\sigma\gamma_{00l}^2$ ($\beta_{00l3}$). 
The corresponding eigenvectors are 
\begin{equation}
\label{eq4.17}
\begin{aligned}
&  \psi^{\beta_{00l1}}(x,y,z)=( 0, 0, 0, 0, \sin l\pi z )^t, &&      
 \psi^{\beta_{00l2}}(x,y,z)=( 0, 0, 0, \sin l\pi z, 0)^t,\\
& \psi^{\beta_{00l \sigma}}_1(x,y,z)= ( \cos l\pi z, 0, 0, 0, 0 )^t, &&
 \psi^{\beta_{00l \sigma}}_2(x,y,z)= ( 0, \cos l\pi z, 0, 0, 0 )^t.                 \end{aligned}
\end{equation}

To analyze the structure of the eigenspaces 
of (\ref{eq4.1}),  
for $k \in \mathbb Z$, $j \in \{0\} \cup \mathbb N$ 
and $l \in \mathbb N$, we define
\begin{align*}
& \phi_{jkl}^1=( \frac{j\alpha_1 l\pi}{\alpha_{jk}^2}
            \cos (j\alpha_1 x+ k\alpha_2 y) \cos l\pi z,
            \frac{k\alpha_2 l\pi}{\alpha_{jk}^2}
            \cos (j\alpha_1 x+ k\alpha_2 y) \cos l\pi z,\\
& \qquad \qquad \qquad \qquad 
            \sin (j\alpha_1 x+k\alpha_2 y) \sin l\pi z,
            0,
            0 )^t ,\\
& \phi_{jkl}^2=( 
            0,
            0,
            0,
            \sin (j\alpha_1 x+ k\alpha_2 y) \sin l\pi z,
            0
            )^t , \\
& \phi_{jkl}^3=( 
            0,
            0,
            0,
            0,
            \sin (j\alpha_1 x+k\alpha_2 y) \sin l\pi z
            )^t,\\
&\phi_{jkl}^4=(
            -\frac{j\alpha_1 l\pi}{\alpha_{jk}^2}
        \sin(j\alpha_1 x+ k\alpha_2 y )\cos l\pi z,
            -\frac{k\alpha_2 l\pi}{\alpha_{jk}^2}
        \sin(j\alpha_1 x+ k\alpha_2 y )\cos l\pi z,\\
& \qquad \qquad \qquad \qquad 
            \cos(j\alpha_1 x+k \alpha_2 y )\sin l\pi z,
            0,
            0 
            )^t,\\
&\phi_{jkl}^5=(
           0,
           0,
           0,
           \cos(j\alpha_1 x+k\alpha_2 y) \sin l\pi z,
           0
           )^t,\\
& \phi_{jkl}^6=(  
            0,
            0,
            0,
            0, 
            \cos (j\alpha_1 x+k\alpha_2 y) \sin l\pi z
            )^t.
\end{align*}

The following lemma follows from (\ref{eq4.1})-(\ref{eq4.7}).

\begin{lemma}
\label{le4.7}
If $j^2+k^2\ne 0$ and  $\beta_{jklq}$ ($q=1,2,3$) 
is a zero of $f_{jkl}$, then we have the followings. 
\begin{enumerate}
\item  The eigenvector corresponding 
to $\beta_{jklq}$ in the complexified space of $H$ is 
\begin{align}
\label{eq4.18}
\psi^{\beta_{jklq}}= e^{i(j\alpha_1 x+k\alpha_2 y)}( 
            \frac{ij \alpha_1 l\pi}{\alpha_{jk}^2}\cos l\pi z,
          &  \frac{ik \alpha_2 l\pi}{\alpha_{jk}^2}\cos l\pi z,   
           \sin l\pi z, \\
          &  A_1(\beta_{jklq})\sin l\pi z, 
           A_2(\beta_{jklq})\sin l\pi z )^t, \nonumber
\end{align}
where 
 $A_1(\beta_{jklq})=\frac{1}{\beta_{jklq}+\gamma_{jkl}^2}, \qquad
 A_2(\beta_{jklq})=\frac{1}{\beta_{jklq}+\tau\gamma_{jkl}^2}$ .

\item If $\beta_{jklq}$ is a real number, the corresponding  eigenvectors are
      given by
\begin{equation}
\label{eq4.19}
\begin{aligned}
& \psi^{\beta_{jklq}}_1=\phi^1_{jkl}+A_1(\beta_{jklq})\phi^2_{jkl}
                +A_2(\beta_{jklq})\phi^3_{jkl}
           \\
& \psi^{\beta_{jklq}}_2=\phi^4_{jkl}+A_1(\beta_{jklq})\phi^5_{jkl}
                +A_2(\beta_{jklq})\phi^6_{jkl}
\end{aligned}
\end{equation}

 \item If $Im(\beta_{jklq})\ne 0$, the generalized eigenvectors corresponding
      to $\beta_{jklq}$ and $\bar{\beta}_{jklq}$ are       
\begin{equation}
\label{eq4.20}
\begin {aligned}
& \psi^{\beta_{jklq}}_1=\phi^1_{jkl}+
                     R_1(\beta_{jklq})\phi^2_{jkl}
                    +R_2(\beta_{jklq})\phi^3_{jkl}+
                    I_1(\beta_{jklq})\phi^5_{jkl}
                    +I_2(\beta_{jklq})\phi^6_{jkl},\\
& \psi^{\beta_{jklq}}_2= -I_1(\beta_{jklq})\phi^2_{jkl}
                         -I_2(\beta_{jklq})\phi^3_{jkl}+\phi^4_{jkl} 
                         +R_1(\beta_{jklq})\phi^5_{jkl}
                         +R_2(\beta_{jklq})\phi^6_{jkl}, \\
& \psi^{\bar{\beta}_{jklq}}_1=\phi^1_{jkl}+
                     R_1(\bar{\beta}_{jklq})\phi^2_{jkl}
                    +R_2(\bar{\beta}_{jklq})\phi^3_{jkl}+
                     I_1(\bar{\beta}_{jklq})\phi^5_{jkl}
                    +I_2(\bar{\beta}_{jklq})\phi^6_{jkl},\\          
& \psi^{\bar{\beta}_{jklq}}_2=  -I_1(\bar{\beta}_{jklq})\phi^2_{jkl}
                        -I_2(\bar{\beta}_{jklq})\phi^3_{jkl}+\phi^4_{jkl} 
                         +R_1(\bar{\beta}_{jklq})\phi^5_{jkl}
                         +R_2(\bar{\beta}_{jklq})\phi^6_{jkl}, \\
\end{aligned}
\end{equation}
where $R_1(\beta)=Re(A_1(\beta))$, $I_1(\beta)=Im(A_1(\beta))$, $R_2(\beta)=
Re(A_2(\beta))$ and $I_2(\beta)=Im(A_2(\beta))$.
\end{enumerate}
\end{lemma}

\begin{definition}
\label{df4.9}
\begin{enumerate}
\item If $j=k=0$, for each $l\in \mathbb N$, we define 
  $ E_{00l}= \text{span}\{\psi^{\beta_{00l1}},
             \psi^{\beta_{00l2}}
               \}$ \, and \, $ E_{00l}^{\sigma}= \text{span}\{ 
              \psi^{\beta_{00l \sigma}}_{1},
              \psi^{\beta_{00l \sigma}}_{2}\}.$
\item For $j^2+k^2 \ne 0 $, we define 
  $ E_{jkl}^1=\text{span}\{\phi_{jkl}^1,
        \phi_{jkl}^2,
        \phi_{jkl}^3\}, \,\,
    E_{jkl}^2=\text{span}\{\phi_{jkl}^4,
      \phi_{jkl}^5,
      \phi_{jkl}^6\}$, \\
     $ E_{jkl}=E_{jkl}^1\oplus E_{jkl}^2$,  and 
     $ E_{jkl}^{\sigma}=\text{span}\{\psi^{\beta_{jkl \sigma}}_1,
        \psi^{\beta_{jkl \sigma}}_2 \}.$

\item For $ (j,k,l)\in \mathbb Z^2 \times \mathbb N$, we define $E_{f_{jkl}}$
      to be the eigenspace spanned by the eigenvectors and the generalized 
      eigenvectors corresponding to the zeros of $f_{jkl}$. 
\end{enumerate}
\end{definition}

\begin{remark}
\label{rm4.10}
{\rm
\begin{enumerate}
\item It is easy to see from the Fourier expansion that 
$\{E_{jkl}\cup E_{jkl}^{\sigma}\}_{j,l=0,
      k=-\infty}^{\infty}$ is a basis of $H_1$.
\item $E_{jkl}$ (resp., $E_{jkl}^{\sigma}$) is orthogonal to $E_{j_1 k_1 l_1}$
      (resp.,$E_{j_1 k_1 l_1}^{\sigma}$) 
      for $(j,k,l)\ne(j_1, k_1, l_1)$, and $E_{jkl}$
      is always orthogonal to $E_{j_1 k_2 l_2}^{\sigma}$.  
\end{enumerate}
}
\end{remark}

The following theorem together with Lemmas~\ref{le4.2} and \ref{le4.4}
complete the analysis of the eigenvalue problem (\ref{eq4.1}).

\begin{theorem}
\label{th4.11}
Under the assumption (\ref{eq4.10}), we have
\begin{enumerate}
\item[1)] $E_{f_{jkl}}=E_{jkl}$ for 
      $j\in\{0\}\cup \mathbb N$, $k\in \mathbb Z$, and $l\in \mathbb N$;
      and
\item[2)] $L_{\lambda \mu}|E_{jkl}$ is strictly negative definite for each
      $(j,k,l)\in \mathbb Z^2 \times \mathbb N$ when $\lambda < \lambda_c$. 

\end{enumerate}
\end{theorem}
\begin{proof}
The first assertion follows by the fact that $E_{jkl}$
is an invariant subspace of $L_{\lambda \eta}$ and 
$f_{jkl}$ is the characteristic polynomial of  
$L_{\lambda \eta} \mid_{E_{jkl}}$ . Assertion 1), 
Lemma~\ref{le4.4} together with the fact that $f_{jkl}$ does 
not have a zero of multiplicity three imply Assertion 2). 

\end{proof}

 We conclude the above analysis as follows..
\begin{enumerate}
\item The eigenvalues of $L_{\lambda \eta}:H_1 \to H$ consist of 
$\{\beta_{jklq},\beta_{jkl \sigma}\}_{j,l=0,k=-\infty}^{\infty}$.
\item The (generalized) eigenvectors of $L_{\lambda \eta}$ 
      form a basis of $H$. 
\item  $-L_{\lambda \eta}$ is a sectorial operator.
\item For the rectangle case, the multiplicity of the first eigenvalue,
      $\beta_{1011}(\lambda)$, is 2, and the corresponding eigenvectors
      are $\psi^{\beta_{1011}}_1$ and $\psi^{\beta_{1011}}_2$.
\item For the square case, the multiplicity of the first eigenvalue,
      $\beta_{1111}(\lambda)$(=$\beta_{1-111}(\lambda)$), is 4, 
      and the corresponding eigenvectors
      are $\psi^{\beta_{1111}}_1$, $\psi^{\beta_{1111}}_2$, 
      $\psi^{\beta_{1-111}}_1$ and $\psi^{\beta_{1-111}}_2$.
\end{enumerate}

 To show that $G$ satisfies condition (\ref{eq3.3}).
 We assume $\frac{3}{4}<\mu<1$, then
 for $\psi\in H_{\mu}\subset H$, by Sobolev's inequality, 
 \begin{eqnarray*}
  |G(\psi)|_{H}^2  \le\int_0^1\int_0^{2\pi/\alpha_2}
   \int_0^{2\pi/\alpha_1}|\psi|^2|\nabla\psi|^2dxdydz                  
        \le |\psi|_{L^{\infty}}^2|\psi|_{H_{1/2}}^2
                 \le C |\psi|_{H_{\mu}}^4. 
 \end{eqnarray*}
   where $C$ is some constant. Hence, $G(\psi)=o(|\psi|_{H_{\mu}})$.

\subsection{Dual Basis}
Since  $E_{jkl}$ is finite dimensional for each 
$(j,k,l)$, there exists a vector $\Psi^{\beta_{jklq}}_{p}\in E_{jkl}$ 
$( q=1,2,3 \,\text{and}\, p=1,2)$
such that
\begin{equation}
\label{eq4.23}
<\Psi^{\beta_{jklq}}_{p},\psi^{\beta_{jklq^*}}_{p^*}>_H \begin{cases}
 & \ne 0 \quad \text{if} \quad (q^*,p^*)=(q,p),\\ 
 & =0  \quad \text{if}  \quad (q^*, p^*)\ne(q,p).
\end{cases}
\end{equation}
 We choose
$$ 
\Psi^{\beta_{jkl \sigma}}_{p}=\psi^{\beta_{jkl \sigma}}_{p}.
$$
Hence, by the orthogonality of $E_{jkl}$ and ${E_{jkl}^{\sigma}}$,
$\{\Psi^{\beta_{jklq}}_p,\Psi^{\beta_{jkl \sigma}}_p\}_{j,l=0,k=-\infty}^{\infty}$ 
form a dual basis of H corresponding to 
$\{E_{jkl},E_{jkl}^{\sigma}\}_{j,l=0,k=-\infty}^{\infty}$ in the sense that
$$
<\Psi^{\beta}_{p},\psi^{\beta^*}_{p^*}>_H \begin{cases}
 & \ne 0 \quad \text{if} \quad (\beta^*,p^*)=(\beta,p),\\ 
 & =0  \quad \text{if}  \quad (\beta^*, p^*)\ne(\beta,p). 
\end{cases}
$$ 

The proof of Lemma~\ref{le4.6} implies that
all but finitely many of $f_{jkl}$ have three distinct real  zeros. 
For such $f_{jkl}$ with  $j^2+k^2 \ne 0$, 
 $\Psi^{\beta_{jklq}}_1$ and
$\Psi^{\beta_{jklq}}_2$ could be chosen as
\begin{equation}
\label{eq4.24}
\begin{aligned}
\begin{cases}
& \Psi^{\beta_{jklq}}_1=\phi_{jkl}^1+C_1(\beta_{jklq})\phi_{jkl}^2
                      +C_2(\beta_{jklq})\phi_{jkl}^3,\\
& \Psi^{\beta_{jklq}}_2=\phi_{jkl}^4+C_1(\beta_{jklq})\phi_{jkl}^5
                      +C_2(\beta_{jklq})\phi_{jkl}^6,
\end{cases}
\end{aligned}
\end{equation}
where
\begin{equation}
\label{eq4.26}
\begin{cases}
& C_1(\beta_{jklq})=\frac{\sigma \lambda}{\beta_{jklq}+\gamma_{jkl}^2}
  =\sigma \lambda A_1(\beta_{jklq}),\\
& C_2(\beta_{jklq})=\frac{-\sigma \eta}{\beta_{jklq}+\tau \gamma_{jkl}^2}
  =-\sigma \eta A_2(\beta_{jklq}).
\end{cases}
\end{equation}

\section { The Proofs of the Main Theorems}
We are now in a position to  reduce equations of (\ref{eq2.4})-
(\ref{eq2.7}) to the center manifold. We would like to fix $\eta<\eta_c$,
and let $\lambda \approx \lambda_c$ be the bifurcation parameter.
 For any $\psi=(U,T,S)\in H$, we have 
\begin{eqnarray*}
\psi=\sum_{k=-\infty}^{\infty}\sum_{j=0,l=1}^{\infty}\sum_{q=1}^{3}
(x_{jklq}\psi_{1}^{\beta_{jklq}}+
     y_{jklq}\psi_{2}^{\beta_{jklq}})
 +\sum_{k=-\infty}^{\infty}\sum_{j,l=0}^{\infty}(x_{jkl\sigma}
 \psi^{\beta_{jkl \sigma}}_1 + y_{jkl \sigma} \psi^{\beta_{jkl \sigma}}_2).  
\end{eqnarray*}
For the square case, we assume $\alpha_1=\alpha_2=\frac{\pi}{2}$.
Hence $\beta_{1111}(\lambda)=\beta_{1-111}(\lambda)$ 
are the first eigenvalues.

 For brevity, we let 
$\beta_0(\lambda)=\beta_{1111}(\lambda)=\beta_{1-111}(\lambda)$,
then the reduced equations are given by

\begin{equation}
\label{eq5.3}
\left\{\begin{aligned}
& \frac{dx_{1111}}{dt}=\beta_0(\lambda)x_{1111}+\frac{1}
  {<\psi_{1}^{\beta_{1111}},\Psi_{1}^{\beta_{1111}}>_H}
  <G( \psi, \psi),\Psi_{1}^{\beta_{1111}})>_H,\\
& \frac{dy_{1111}}{dt}=\beta_0(\lambda)y_{1111}+\frac{1}
  {<\psi_{2}^{\beta_{1111}},\Psi_{2}^{\beta_{1111}}>_H}
  <G( \psi, \psi),\Psi_{2}^{\beta_{1111}})>_H,\\
& \frac{dx_{1-111}}{dt}=\beta_0(\lambda)x_{1-111}+\frac{1}
  {<\psi_{1}^{\beta_{1-111}},\Psi_{1}^{\beta_{1-111}}>_H}
  <G( \psi, \psi),\Psi_{1}^{\beta_{1-111}})>_H,\\
& \frac{dy_{1-111}}{dt}=\beta_0(\lambda)y_{1-111}+\frac{1}
  {<\psi_{2}^{\beta_{1-111}},\Psi_{2}^{\beta_{1-111}}>_H}
  <G( \psi, \psi),\Psi_{2}^{\beta_{1-111}})>_H.
\end{aligned}
\right.
\end{equation} 

Here for $\psi_1=(U_1,T_1,S_1)$, $\psi_2=(U_2,T_2,S_2)$ and 
$\psi_3=(U_3,T_3,S_3)$,
\begin{align*} 
<G(\psi_1,\psi_2),\psi_3>_H= & -\int_{0}^{1}
    \int_{0}^{2\pi/\alpha_2} 
    \int_{0}^{2\pi/\alpha_1}[<(U_1\cdot\nabla)U_2,
         U_3>_{\mathbb R^3} \\
  & +(U_1\cdot\nabla)T_2 T_3 
        +(U_1\cdot\nabla)S_2 S_3]dxdydz. 
\end{align*}
Let the center manifold function be denoted by 
\begin{equation}
\label{eq5.6}
\Phi=\sum_{\beta \ne \beta_{1111}, \beta_{1-111}}( \Phi_{1}^{\beta}(x_{1111},
      y_{1111},x_{1-111},y_{1-111})\psi_{1}^{\beta}+
     \Phi_{2}^{\beta}(x_{111},y_{111},x_{1-111},y_{1-111})\psi_{2}^{\beta}).
\end{equation}
 Note that for any $\psi_i \in H_1$ ($i=$ 1, 2, 3), $<G(\psi_1,\psi_2),\psi_2>_H=0$, $<G(\psi_1,\psi_2),\psi_3>_H=-<G(\psi_1,\psi_3),\psi_2>_H$,
and for any $\psi_{i}\in E_{jkl}$ $(i=1,2,3)$, $<G(\psi_1,\psi_2),\psi_3>_H=0$.
Applying Theorem 3.8 in \cite{b-book}, we obtain
\begin{align}
& \Phi^{\beta_{0021}}=\frac{A_2\pi}{2\beta_{0021}}[x_{1111}^2
  +y_{1111}^2+x_{1-111}^2+y_{1-111}^2]+o(2),
\nonumber \\
& \Phi^{\beta_{0022}}=\frac{A_1\pi}{2\beta_{0022}}[x_{1111}^2
  +y_{1111}^2+x_{1-111}^2+y_{1-111}^2]+o(2),
\nonumber\\
&\Phi_1^{\beta_{202q}}=B(\beta_{202q})(x_{1111}y_{1-111}+x_{1-111}y_{1111})
+o(2),
\nonumber\\
&\Phi_2^{\beta_{202q}}=B(\beta_{202q})(-x_{1111}x_{1-111}+y_{1111}y_{1-111})
+o(2), 
\nonumber\\
& \Phi_1^{\beta_{022q}}=B(\beta_{022q})(x_{1111}y_{1-111}-x_{1-111}y_{1111})
+o(2),
\nonumber\\
& \Phi_2^{\beta_{022q}}=B(\beta_{022q})(x_{1111}x_{1-111}+y_{1111}y_{1-111})
+o(2), 
\nonumber
\end{align}
\begin{align}
\label{eq5.7}
& \Phi(x_{1111},y_{1111},x_{1-111},y_{1-111})=
\Phi^{\beta_{0021}}\psi^{\beta_{0021}}+\Phi^{\beta_{0022}}\psi^{\beta_{0022}}\\
& +\sum_{q=1}^{3}(\Phi_1^{\beta_{202q}}\psi_1^{\beta_{202q}}
               +\Phi_2^{\beta_{202q}}\psi_2^{\beta_{202q}}
               +\Phi_1^{\beta_{022q}}\psi_1^{\beta_{022q}}
               +\Phi_2^{\beta_{022q}}\psi_2^{\beta_{022q}})
+o(2),
\nonumber
\end{align}
where
\begin{equation}
\label{eq5.8} 
\begin{aligned}
  B(\beta)=& \frac{\pi(3+A_1(\beta_0)C_1(\beta)+A_2(\beta_0)C_2(\beta))}{2
   \beta(5+A_1(\beta)C_1(\beta)+A_2(\beta)C_2(\beta))},\\
  C_1(\beta)=&\frac{\sigma \lambda}{
   (\beta+\gamma_{202}^2)}, \qquad
 C_2(\beta)=\frac{\sigma \eta}{
   (\beta+\tau \gamma_{202}^2)},\\
 o(2)=& o(x_{1111}^2+y_{1111}^2+x_{1-111}^2+y_{1-111}^2)\\
     & +O(|\beta_0(\lambda)| \cdot (x_{1111}^2+y_{1111}^2
     +x_{1-111}^2+y_{1-111}^2)).
\end{aligned}
\end{equation}
Hereafter, we make the following convention,
\begin{align*}
 o(3)=& o((x_{1111}^2+y_{1111}^2+x_{1-111}^2+y_{1-111}^2)^{3/2})\\
     & +O(|\beta_0(\lambda)| \cdot (x_{1111}^2+y_{1111}^2
     +x_{1-111}^2+y_{1-111}^2)^{3/2}),  \\
 o(4)=& o((x_{1111}^2+y_{1111}^2+x_{1-111}^2+y_{1-111}^2)^{2})\\
     & +O(|\beta_0(\lambda)| \cdot (x_{1111}^2+y_{1111}^2
     +x_{1-111}^2+y_{1-111}^2)^{2}). 
\end{align*} 
By (\ref{eq5.7}) and the fact that $\beta_{022q}=\beta_{202q}$, we obtain
\begin{align*}
 <G(\psi,\psi),\Psi^{\beta_{1111}}_1>_H =&<G(\psi_1^{\beta_{1111}},
      \Phi),\Psi_1^{\beta_{1111}}>_H
       +<G(\psi_2^{\beta_{1111}},\Phi),\Psi_1^{\beta_{1111}}>_H\\
      & +<G(\psi_1^{\beta_{1-111}},\Phi),\Psi_1^{\beta_{1111}}>_H
       +<G(\psi_2^{\beta_{1-111}}),\Phi),\Psi_1^{\beta_{1111}}>_H
       +o(3)\\
     =& -<G(\psi_1^{\beta_{1111}},\Psi_1^{\beta_{1111}}),\Phi>_H
       -<G(\psi_2^{\beta_{1111}},\Psi_1^{\beta_{1111}}),\Phi>_H\\
      & -<G(\psi_1^{\beta_{1-111}},\Psi_1^{\beta_{1111}}),\Phi>_H
      -<G(\psi_2^{\beta_{1-111}}),\Psi_1^{\beta_{1111}}),\Phi>_H
       +o(3)\\
     =&2\pi^2 (\frac{A_1 C_1}
                {\beta_{0022}}+\frac{A_2 C_2}{\beta_{0021}})
                (x_{1111}^2+y_{1111}^2+x_{1-111}^2+y_{1-111}^2)x_{1111}\\
         &+2\sum_{q=1}^{3}B^*(\beta_{202q})B(\beta_{202q})
          (x_{1-111}^2+y_{1-111}^2)x_{1111}+o(3),     
 \end{align*}
where
\begin{equation}
\label{eq5.9}
 B^*(\beta)=\pi(3+A_1(\beta)C_1(\beta_0)+A_2(\beta)C_2(\beta_0)).
\end{equation}

Similarly, we derive that
\begin{align*}
 <G(\psi,\psi),\Psi^{\beta_{1111}}_2>_H=&2\pi^2 (\frac{A_1 C_1}
                {\beta_{0022}}+\frac{A_2 C_2}{\beta_{0021}})
                (x_{1111}^2+y_{1111}^2+x_{1-111}^2+y_{1-111}^2)y_{1111}\\
         &+2\sum_{q=1}^{3}B^*(\beta_{202q})B(\beta_{202q})
              (x_{1-111}^2+y_{1-111}^2)y_{1111}+o(3),
\end{align*}
\begin{align*}
 <G(\psi,\psi),\Psi^{\beta_{1-111}}_1>_H=&2\pi^2 (\frac{A_1 C_1}
                {\beta_{0022}}+\frac{A_2 C_2}{\beta_{0021}})
                (x_{1111}^2+y_{1111}^2+x_{1-111}^2+y_{1-111}^2)x_{1-111}\\
         &+2\sum_{q=1}^{3}B^*(\beta_{202q})B(\beta_{202q})
              (x_{1111}^2+y_{1111}^2)x_{1-111}+o(3),
\end{align*}
\begin{align*}
 <G(\psi,\psi),\Psi^{\beta_{1-111}}_2>_H=&2\pi^2 (\frac{A_1 C_1}
                {\beta_{0022}}+\frac{A_2 C_2}{\beta_{0021}})
                (x_{1111}^2+y_{1111}^2+x_{1-111}^2+y_{1-111}^2)y_{1-111}\\
         &+2\sum_{q=1}^{3}B^*(\beta_{202q})B(\beta_{202q})
              (x_{1111}^2+y_{1111}^2)y_{1-111}+o(3).
\end{align*}

Therefore, (\ref{eq5.3}) can be rewritten as
\begin{equation}
\label{eq5.10}
\left\{\begin{aligned}
& \frac{dx_{1111}}{dt}=\beta_0(\lambda)x_{1111}+
   \delta_1 E x_{1111}+
   \delta_2(x_{1-111}^2+y_{1-111}^2)x_{1111}
   +o(3),\\
& \frac{dy_{1111}}{dt}=\beta_0(\lambda)y_{1111}+
   \delta_1 E y_{1111}+
   \delta_2(x_{1-111}^2+y_{1-111}^2)y_{1111}
   +o(3),\\
& \frac{dx_{1-111}}{dt}=\beta_0(\lambda)x_{1-111}+
   \delta_1 E x_{1-111}+
   \delta_2(x_{1111}^2+y_{1111}^2)x_{1-111}
   +o(3),\\
& \frac{dy_{1-111}}{dt}=\beta_0(\lambda)y_{1-111}+
   \delta_1 E y_{1-111}+
   \delta_2(x_{1111}^2+y_{1111}^2)y_{1-111}
   +o(3),\\
\end{aligned}
\right.
\end{equation} 
where $E=(x_{1111}^2+y_{1111}^2+x_{1-111}^2+y_{1-111}^2),$
\begin{align}
\label{eq5.11}
\delta_1=&\delta_1(\lambda,\eta)=2\pi^2 (\frac{A_1 C_1}
                {\beta_{0022}}+\frac{A_2 C_2}{\beta_{0021}})/
               4(3+A_1C_1+A_2C_2)\\
        =&-\frac{(A_1C_1+\tau^{-1}A_2C_2)}{8(3+A_1C_1+A_2C_2)},
   \nonumber\\
\label{eq5.12}
 \delta_2=&\delta_2(\lambda,\eta)
   =2\sum_{q=1}^{3}B^*(\beta_{202q})B(\beta_{202q})/ 4(3+A_1C_1+A_2C_2).
\end{align}
The energy estimate of (\ref{eq5.10}) is given by
\begin{equation}
\label{eq5.13}
\frac{1}{2}\frac{dE}{dt}
  =\beta_0(\lambda)E+\delta_1 E^2 
   +2 \delta_2 (x_{1111}^2+y_{1111}^2)(x_{1-111}^2+y_{1-111}^2)+o(4).
\end{equation}

If the quadratic form
\begin{equation}
\label{eq5.14}
\delta_1 E^2 
   +2 \delta_2 (x_{1111}^2+y_{1111}^2)(x_{1-111}^2+y_{1-111}^2),
\end{equation}
is negative definite at $\lambda=\lambda_c$, we conclude that $(U,T,S)=0$
is a locally asymptotically stable equilibrium point of 
(\ref{eq2.4})-(\ref{eq2.7}).
We then obtain Theorem~\ref{th2.6} by Theorem~\ref{th3.6} directly.
And if the quadratic form (\ref{eq5.14}) is positive definite, 
we obtain the subcritical  bifurcation. 
This completes the proofs of Theorem~\ref{th2.6} and
Theorem~\ref{th2.7}.

For the rectangle case, the first eigenspace is of dimension two. The reduced
equations on the center manifold  are given by
\begin{equation}
\label{eq5.15}
\left\{\begin{aligned}
& \frac{dx_{1011}}{dt}=\beta_{1011}(\lambda)x_{1011}
                  +\delta_1 (x_{1011}^2+y_{1011}^2)x_{1011}+o(3),\\
& \frac{dy_{1011}}{dt}=\beta_{1011}(\lambda)y_{1011}
                  +\delta_1 (x_{1011}^2+y_{1011}^2)y_{1011}+o(3).
\end{aligned}\right.
\end{equation}
The rest part of the proofs of Theorem~\ref{th2.3}, Theorem~\ref{th2.4}  
and Theorem~\ref{th2.5} 
is similar to the 2D case in \cite{hmw}. In this case, we can use 
Theorem 5.10 in \cite{amsbook}  or invariant sphere theorem in \cite{field} 
to conclude the $S^1$ structure of the bifurcating solution.
 
\begin{remark}
{\rm  
\begin{enumerate}
\item As we have already proved in \cite{hmw}, we know that  
\begin{equation}
\label{eq5.16}
\delta_1\begin{cases}
& <0  \quad  \text{if}\quad  \eta<\eta_{c_1},\\
& >0  \quad  \text{if}\quad  \eta>\eta_{c_1}.
\end{cases}
\end{equation}
\item When $\sigma$, $\tau$, $\lambda$ and $\eta$ are given, $\delta_1$ and
      $\delta_2$ can be evaluated numerically.
\end{enumerate}
}
\end{remark} 

\subsection*{Acknowledgements}
The authors would like to thank the referee's insightful comments 
and suggestions.  The work was supported in part by the
Office of Naval Research,  by the National Science Foundation, 
and by the National Science Foundation
of China.

\bibliography{3zaa}

\begin{thebibliography}{10}

\bibitem{pg}
{\sc P.~G. Baines and A.~Gill}, {\em On thermohaline convection with linear
  gradients}, J. Fluid Mech., 37 (1969), pp.~289--306.

\bibitem{busse}
{\sc F.~H. Busse}, {\em On the stability of two-dimensional convection in a
  layer heated from below}, J. Math.\& Phys., 46 (1976), pp.~140--143.

\bibitem{bg1}
{\sc E.~Buzano and M.~Golubitsky}, {\em Bifurcation on the hexagonal lattice
  and the planar {B}\'enard problem}, Philos. Trans. Roy. Soc. London Ser. A,
  308 (1983), pp.~617--667.

\bibitem{field}
{\sc M.~Field}, {\em Equivariant bifurcation theory and symmetry breaking}, J.
  Dynam. Differential Equations, 1 (1989), pp.~369--421.

\bibitem{fmt}
{\sc C.~Foias, O.~Manley, and R.~Temam}, {\em Attractors for the {B}\'enard
  problem: existence and physical bounds on their fractal dimension}, Nonlinear
  Anal., 11 (1987), pp.~939--967.

\bibitem{bg2}
{\sc M.~Golubitsky, I.~Stewart, and D.~G. Schaeffer}, {\em Singularities and
  groups in bifurcation theory. {V}ol. {II}}, vol.~69 of Applied Mathematical
  Sciences, Springer-Verlag, New York, 1988.

\bibitem{henry}
{\sc D.~Henry}, {\em Geometric theory of semilinear parabolic equations},
  vol.~840 of Lecture Notes in Mathematics, Springer-Verlag, Berlin, 1981.

\bibitem{hmw}
{\sc C.-H. Hsia, T.~Ma, and S.~Wang}, {\em Bifurcation and stability of
  two-dimensional double-diffusive convection}, submitted,  (2006).

\bibitem{jevons}
{\sc W.~S. Jevons}, {\em On the cirrous form of cloud}, London, Edinburgh,
  Dublin Philos. Mag. J. Sci. Ser. 4, 14 (1857), pp.~22--35.

\bibitem{LTW92b}
{\sc J.~L. Lions, R.~Temam, and S.~Wang}, {\em On the equations of large-scale
  ocean}, Nonlinearity, 5 (1992), pp.~1007--1053.

\bibitem{b-book}
{\sc T.~Ma and S.~Wang}, {\em Bifurcation Theory and Applications}, vol.~53 of
  World Scientific Series on Nonlinear Science, Series A, World Scientific,
  2005.

\bibitem{mw-db1}
\leavevmode\vrule height 2pt depth -1.6pt width 23pt, {\em Dynamic bifurcation
  of nonlinear evolution equations and applications}, Chinese Annals of
  Mathematics, 26:2 (2005), pp.~185--206.

\bibitem{amsbook}
\leavevmode\vrule height 2pt depth -1.6pt width 23pt, {\em Geometric Theory of
  Incompressible Flows with Applications to Fluid Dynamics}, vol.~119 of
  Mathematical Surveys and Monographs, American Mathematical Society,
  Providence, RI, 2005.

\bibitem{wn}
{\sc W.~Nagata and J.~W. Thomas}, {\em Bifurcation in double-diffusive systems
  i. equilibrium solutions}, SJMA, 17:1 (1986), pp.~91--113.

\bibitem{sattinger}
{\sc D.~H. Sattinger}, {\em Group representation theory, bifurcation theory and
  pattern formation}, J. Funct. Anal., 28 (1978), pp.~58--101.

\bibitem{stern}
{\sc M.~E. Stern}, {\em The ``salt fountain'' and thermohaline convection},
  Tellus, 12 (1960), pp.~172--175.

\bibitem{stommel}
{\sc H.~Stommel, A.~Arons, and D.~Blanchard}, {\em An oceanographical curiosity
  : the perpetual salt fountain}, Deep-Sea Res., 3 (1956), pp.~152--153.

\bibitem{temam}
{\sc R.~Temam}, {\em Infinite-dimensional dynamical systems in mechanics and
  physics}, vol.~68 of Applied Mathematical Sciences, Springer-Verlag, New
  York, second~ed., 1997.

\bibitem{gv}
{\sc G.~Veronis}, {\em On finite amplitude instability in the thermohaline
  convection}, J. Marine Res., 23 (1965), pp.~1--17.

\end{thebibliography}

\end{document}